\documentclass[useAMS]{mn2e}
\usepackage{times}
\usepackage{graphicx}
      
\title[Blazar distance indications from {\it Fermi} and TeV observations]
{Constraining blazar distances with combined 
 {\it Fermi} and TeV data: an empirical approach}

\author[Prandini et al.]
{E. Prandini$^1$\thanks{E--mail: prandini@pd.infn.it}, G. Bonnoli$^2$,
L. Maraschi$^3$, M. Mariotti$^1$, F. Tavecchio$^2$ \\
$^1$ Dipartimento di Fisica, Padova University and INFN Sez. di Padova, via Marzolo 8, I--35131 Padova, Italy\\
$^2$ INAF -- Osservatorio Astronomico di Brera, via E. Bianchi 46, I--23807 Merate, Italy\\
$^3$ INAF -- Osservatorio Astronomico di Brera, via Brera 28, I--20100 Milano, Italy\\
}
\begin{document}

\pagerange{\pageref{firstpage}--\pageref{lastpage}} \pubyear{}
\maketitle

\label{firstpage}

\begin{abstract} 
We discuss a method to constrain
the distance of blazars with unknown redshift using combined 
observations in the 
GeV and TeV regimes. We assume that 
the VHE spectrum corrected for the absorption through the interaction with
the Extragalactic Background Light can not be harder than the spectrum 
in the {\it Fermi}/LAT band. Starting from the observed VHE spectral data
we derive the EBL-corrected spectra as a function of the redshift $z$ and fit them with 
power laws to be compared with power law fits to the LAT data. 
We apply the method to all TeV blazars detected by LAT 
with known distance 
and derive an empirical law describing the relation between
the upper limits and the true redshifts that can be used 
to estimate the distance of unknown
redshift blazars. Using different EBL models leads to 
systematic changes in the derived upper limits.
Finally, we use this relation to infer the distance of the 
unknown redshift blazar PKS~1424+240.
\end{abstract}

\begin{keywords} 
 galaxies: distances and redshifts - gamma-rays: observations - radiation mechanisms: non--thermal 
\end{keywords}

\section{Introduction}
The extragalactic TeV sky catalogue ($E>100$ GeV),
counts nowadays 35 objects\footnote{for an
updated list see: http://www.mppmu.mpg.de/$\sim$rwagner/sources/}. 
Many of these sources have recently been detected also at
GeV energies  by the {\it Fermi} satellite~(Abdo~et~al. 2009), allowing 
for the first time a quasi-continuous coverage of the spectral shape 
of extragalactic VHE emitters over more 
than 4 decades of energy. 
Except for two starburst galaxies and 
two radiogalaxies, 
all the others are blazars, radio-loud active
galactic nuclei with a relativistic jet closely oriented toward 
the Earth, as described in Urry $\&$ Padovani~(1995).
The apparent luminosity of the non-thermal radiation emitted by the jet 
is then largely enhanced by relativistic beaming and dominates the observed high energy
emission.
Typically, the spectral energy distribution (SEDs) 
emitted from these objects, extending from
radio waves to gamma-ray frequencies, is composed of two broad humps. 
In the case of TeV detected blazars, the first component usually peaks in the 
UV-X-ray band, and the second peak is located at GeV-TeV energies.
The first component is identified as electron synchrotron radiation, whilst the
second component is widely attributed to inverse Compton scattering of ambient 
photons by the same synchrotron emitting electrons. Relativistic electrons are 
accelerated within a region in
bulk relativistic motion along the jet~(e.g. Tavecchio~et~al.~1998). 

VHE photons emitted by cosmological sources 
are effectively absorbed, through the pair production process, 
$\gamma \gamma \rightarrow e^{+-}$, by the interaction 
with the so-called Extragalactic Background Light (EBL)
 (Stecker, de Jager $\&$ Salamon 1992).
EBL is composed of stellar light emitted and partially 
reprocessed by dust throughout the entire history of 
cosmic evolution. The expected EBL spectrum  
is composed by two bumps at near-infrared and far-infrared
wavelengths~(Hauser~$\&$~Dwek~2001). Direct measurement of the EBL 
has proved to be a difficult task, primarily
due to the zodiacal light that forms a bright
foreground which is difficult to suppress.
Due to the lack of direct EBL knowledge, many models
have been elaborated in the last
years~(Stecker, Malkan $\&$ Scully~2006; Franceschini, Rodighiero $\&$ Vaccari~2008; 
Gilmore~et~al.~2009; Kneiske $\&$ Dole~2010).
Moreover, for some blazars the derivation of the
intrinsic spectrum is also difficult due to the uncertainty or lack of a
redshift measurement.
In particular a direct spectroscopic measure of the redshift 
is often difficult in BL Lac 
objects, which are characterized by extremely weak emission lines 
(equivalent width $<$ 5 $\AA$).
\begin{center}
  \begin{table*}
    {\small
      \centering
      \begin{tabular}{llcccccc}
        \hline
        Source Name     & $z[real]$ &{\it Fermi}/LAT   &VHE              & $z^*$      & $z^*$         & $z^*$           & $z[rec]$  \\         
        &        & slope       &slope            &low EBL model  & mean EBL model    &high EBL model&mean EBL model \\ 
        \hline
        \hline
        Mkn 421          & 0.030 & 1.78$\pm$0.03  & 2.3$\pm$ 0.1$^{(1)}$  & 0.101$^{+0.021}_{-0.022}$    & 0.078$^{+0.016}_{-0.018}$  & 0.054$^{+0.012}_{-0.012}$ &  0.009$^{+0.012}_{-0.014}$ \\
        \hline
        Mkn 501          & 0.034 & 1.73$\pm$0.06  & 2.3$\pm$ 0.1$^{(2)}$  & 0.122$^{+0.025}_{-0.024}$    & 0.096$^{+0.018}_{-0.018}$  & 0.067$^{+0.013}_{-0.014}$ &  0.029$^{+0.013}_{-0.013}$ \\
        \hline
        1ES 2344$+$514   & 0.044 & 1.76$\pm$0.27 &2.9$\pm$0.1$^{(3)}$  & 0.248$^{+0.080}_{-0.077}$    & 0.196$^{+0.056}_{-0.060}$  & 0.139$^{+0.043}_{-0.045}$ &   0.105$^{+0.041}_{-0.043}$ \\
        \hline
        Mkn 180          & 0.045 & 1.91$\pm$0.18 &3.3$\pm$0.7$^{(4)}$ & 0.248$^{+0.146}_{-0.150}$    & 0.196$^{+0.116}_{-0.118}$  & 0.147$^{+0.091}_{-0.089}$ &  0.104$^{+0.085}_{-0.086}$ \\
        \hline
        1ES 1959$+$650   & 0.047 & 1.99$\pm$0.09  &2.6$\pm$ 0.2$^{(5)}$  & 0.111$^{+0.055}_{-0.049}$    & 0.086$^{+0.040}_{-0.040}$  & 0.058$^{+0.028}_{-0.017}$ &  0.022$^{+0.029}_{-0.029}$ \\
        \hline
        BL Lacertae      & 0.069 & 2.43$\pm$0.10  &3.6$\pm$0.5$^{(6)}$ & 0.299$^{+0.152}_{-0.162}$    & 0.234$^{+0.116}_{-0.116}$  & 0.172$^{+0.085}_{-0.086}$ &  0.132$^{+0.085}_{-0.085}$ \\
        \hline
        PKS 2005$-$489   & 0.071 & 1.91$\pm$0.09 & 3.2$\pm$0.2$^{(7)}$ & 0.240$^{+0.049}_{-0.047}$    & 0.186$^{+0.036}_{-0.036}$  & 0.129$^{+0.028}_{-0.027}$  &   0.098$^{+0.026}_{-0.026}$ \\
        \hline
        W Comae          & 0.102 & 2.02$\pm$0.06 & 3.7$\pm$0.2$^{(8)}$& 0.298$^{+0.065}_{-0.055}$    & 0.234$^{+0.046}_{-0.046}$  & 0.179$^{+0.036}_{-0.033}$ &   0.133$^{+0.034}_{-0.034}$ \\  
        \hline
        PKS 2155$-$304   & 0.116 & 1.87$\pm$0.03  & 3.4$\pm$0.1$^{(9)}$ & 0.281$^{+0.018}_{-0.017}$    & 0.220$^{+0.014}_{-0.014}$  & 0.162$^{+0.011}_{-0.011}$ &   0.126$^{+0.010}_{-0.010}$ \\
        \hline
        1ES 0806$+$524   & 0.138 & 2.04$\pm$0.14  & 3.6$\pm$1.0$^{(10)}$& 0.281$^{+0.180}_{-0.186}$    & 0.226$^{+0.138}_{-0.152}$  & 0.181$^{+0.107}_{-0.123}$ &    0.126$^{+0.101}_{-0.111}$\\
        \hline
        1ES 1218$+$304   & 0.182 & 1.63$\pm$0.12  &3.1$\pm$0.3$^{(11)}$& 0.264$^{+0.107}_{-0.103}$    & 0.212$^{+0.080}_{-0.082}$  & 0.169$^{+0.061}_{-0.067}$ &    0.114$^{+0.058}_{-0.059}$  \\ 
        \hline
        1ES 1011$+$496   & 0.212 & 1.82$\pm$0.05 &4.0$\pm$0.5$^{(12)}$& 0.667$^{+0.188}_{-0.193}$    & 0.490$^{+0.118}_{-0.124}$  & 0.348$^{+0.112}_{-0.090}$ &    0.323$^{+0.087}_{-0.092}$ \\ 
        \hline
        S5 0716$+$714    & 0.310$^{*,a}$ & 2.16$\pm$0.04 &3.4$\pm$0.5$^{(13)}$& 0.264$^{+0.107}_{-0.117}$    & 0.210$^{+0.086}_{-0.090}$  & 0.157$^{+0.064}_{-0.068}$ &  0.114$^{+0.063}_{-0.066}$\\
        \hline
        PG 1553+113      & 0.400$^b$ & 1.69$\pm$0.04 & 4.1$\pm$0.2$^{(14)}$& 0.779$^{+0.075}_{-0.064}$    & 0.568$^{+0.046}_{-0.046}$  & 0.395$^{+0.031}_{-0.030}$ &    0.338$^{+0.029}_{-0.029}$  \\ 
        \hline
        3C66A            & 0.444$^*$ & 1.93$\pm$0.04  & 4.1$\pm$0.4$^{(15)}$& 0.446$^{+0.076}_{-0.069}$    & 0.344$^{+0.050}_{-0.048}$  & 0.265$^{+0.038}_{-0.039}$ &  0.213$^{+0.037}_{-0.035}$\\
        \hline
        3C279    & 0.536 & 2.34$\pm$0.03  & 4.1$\pm$0.7$^{(16)}$&  1.095$^{+1.066}_{-1.066}$& 0.746$^{+0.716}_{-0.716}$ & 0.440$^{+0.422}_{-0.422}$  &   0.507$^{+0.524}_{-0.524}$  \\ 
        \hline
      \end{tabular}
      \caption
      {TeV blazars used in this study. In the first column 
        the list of sources, their redshift (second column), their 
        {\it Fermi}/LAT slope (third column), the VHE slope of the observed spectrum fit 
        (fourth column). The next 3 columns show the redshift values obtained by 
        de-absorbing the VHE spectra until the slope is the one observed by LAT, 
        using three different EBL models, while the last column lists the corresponding 
        reconstructed redshift of each source obtained by using $z^*$ and the fits parameters,
        as described in the text. $^*$: uncertain. $^a$: from Nilsson et al. (2008). 
        $^b$:private communication with C.~W.~Danforth. 
        1: Acciari et al. (2009d); 2: Albert et al. (2007d); 
        3: Albert et al. (2007a); 4: Albert et al. (2006); 5: Tagliaferri et al. (2008); 
        6: Albert et al. (2007b); 7: Acero et al. (2010); 8: Acciari et al. (2009e);
        9: Aharonian et al. (2005);
        10: Acciari et al (2009a); 11 Acciari et al. (2009c);
        12: Albert et al. (2007c); 13: Anderhub et al. (2009);
        14: Prandini et al. (2009); 15: Acciari et al. (2009b); 16: Albert et al. (2008)}
      \label{table_values}
    }
  \end{table*}
\end{center}
In this paper we discuss a method to derive upper limits on the redshift of 
a source based on the comparison between the spectral index at GeV energies
as measured by LAT (unaffected by the cosmological absorption up to redshifts
far beyond those of interest here) 
and the deabsorbed TeV spectrum. Basically, for larger distances
 the deabsorbed spectrum becomes harder. A solid 
upper limit to the redshift can be inferred deriving the redshift at which 
the slope of the deabsorbed spectrum coincides with that measured by LAT.
Our approach can be considered complementary to those used 
by Stecker $\&$ Scully (2010) and Georganopoulos,~Finke $\&$ Reyes (2010)
(see also Abdo~et~al.~2009), 
where  the comparison of the spectral slopes
at GeV and TeV energies of blazars at known 
distances is used to derive limits on the EBL. 
Starting from the derived limits, we find a 
simple law relating these values to real redshift, that can 
be used to guess the distance of unknown redshift blazars.

We assume a cosmology with $h=0.72$, $\Omega_M=0.3$ and $\Omega_\Lambda=0.7$.

\section{Blazars spectral break}

We consider the blazar sample containing all the extragalactic 
TeV emitters located at redshift larger than $z=0.01$, 
detected  by LAT after 5.5  months of data taking 
as reported in Abdo~et~al.~(2009).
The photon flux emitted by a blazar in both GeV and TeV regimes
 can be usually well approximated with power laws,
of the form $dN/dE = f_0 (E/E_0)^{-\Gamma}$, where $\Gamma$
is the power law index.

\begin{figure}
  \centering
  \vspace*{-1.5 truecm}
  \includegraphics[width=3.5in]{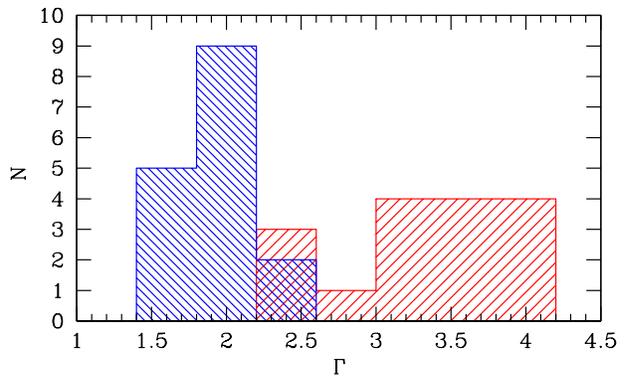}
  \vspace*{-2.7 truecm}
  \caption{Spectral indices distributions of blazars listed in 
    table~\ref{table_values} as measured by
    {\it Fermi}/LAT (blue) and Cherenkov instruments (red).}
  \label{slope_histo}
\end{figure}

Fig.~\ref{slope_histo} represents the comparison between the power law indices,
listed in Table~\ref{table_values},
obtained by fitting the photon spectra of sixteen sources
measured by {\it Fermi}/LAT in the GeV regime
and the slopes  
in the TeV regime measured with the new generation of 
Cherenkov instruments (H.E.S.S., Magic and Veritas).
The spectral slopes $\Gamma$  
in the $0.2-300$~GeV energy range
distribute from $1.63$ to $2.43$, with a peak around 2, while
the VHE spectral slopes show a wider distribution,
ranging from $2.28$ to $4.12$. 

\begin{center}
 \begin{figure*}
 {\small
   \includegraphics[width=7.2in]{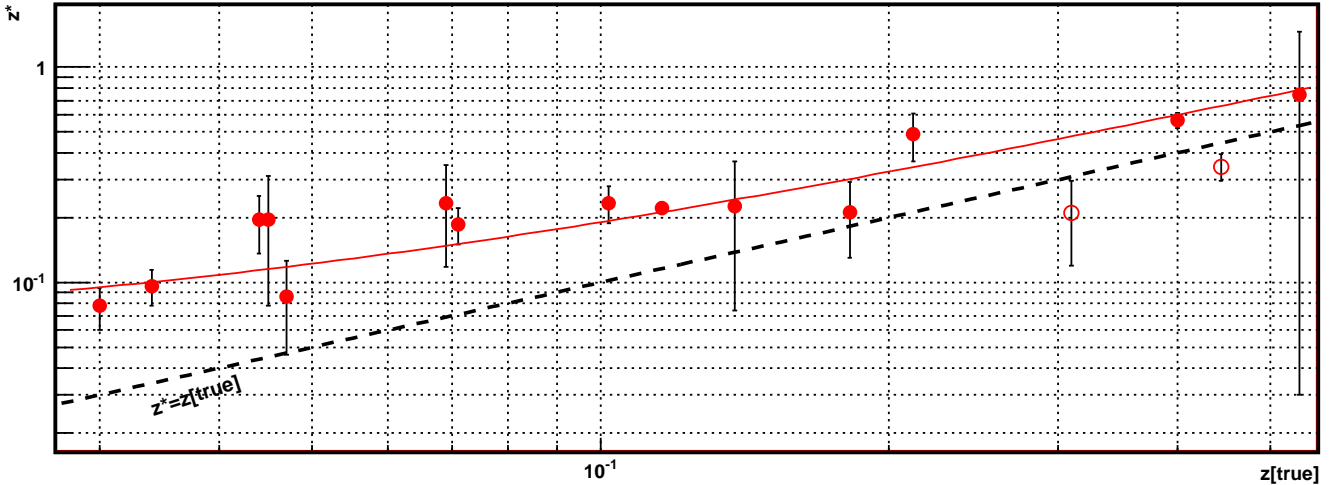}
   \caption{True redshifts vs $z^*$ derived with the procedure described in the text for 
     the Franceschini~et~al.~(2008) EBL model. The open points were not used in 
     the fit calculation (red line) since their redshift is uncertain (sources 3C~66A and S5~0716+714). 
     The dashed line is the bisector: only for the sources 3C~66A and S5~0716+714
     the limits on the redshift estimated in this work are below the true (even if uncertain) redshifts.}
   \label{correlationplot}}
\end{figure*}
\end{center}
 
The systematic difference between the  two distributions is primarily due to an intrinsic 
break in the spectrum emitted by the source.
In fact the peak of the high-energy component in the SED of TeV blazars 
is commonly
located between GeV and thousands of GeV (e.g. Tavecchio et al. 2010): LAT
observes mainly the photons of energy below the energy of the IC peak,  
in the hard portion of the spectrum, while Cherenkov instruments probe the
steeper part of the peak (e.g. Aharonian et al. 2009). 

A second effect influencing the distribution
of TeV slopes is the interaction of VHE photons with EBL
and the consequent reduction of the flux which depends on the distance. 
This dependence is likely responsible for the observed spread of the
TeV spectral indices 
not present in the well peaked GeV indices distribution.

Quantitatively, the effect of the interaction of VHE photons with EBL
is an exponential attenuation of the flux 
by a factor $\tau (E,z)$, where $\tau$
is the optical depth, function of 
both photon energy and source redshift. Thus, the observed
differential energy spectrum from a blazar 
is related to the emitted one according 
to $F_{\rm obs}(E)=e^{-\tau(E)} F_{\rm em}(E)$.
In principle it is possible to derive 
the emitted (or intrinsic) spectrum by deabsorbing
the observed spectrum. This procedure depends
on the absorption coefficient $\tau(E,z)$ and the 
redshift $z$ of the source.
Vice versa, if the intrinsic source spectrum is known, 
given the absorption coefficient  $\tau$,  
the redshift $z$ can be estimated comparing 
the absorbed spectrum with the observed one. 

Here, we use the second approach, developing an empirical method to 
estimate a safe upper limit to the source distance based
on the reasonable assumption that the intrinsic spectrum at TeV
energies cannot be harder than that in the adjacent GeV band. 
Indeed, from the brightest objects studied at both GeV and TeV energies it appears 
that the SED is continuous with a broad peak not requiring 
additional spectral components (e.g.~Aharonian~et~al.~2009). 
Hence, a natural assumption is to require that the
slope measured in the GeV energy range is a 
limit value for the power law index
of the deabsorbed TeV spectrum. This condition, satisfied
when the IC peak maximum extends beyond the VHE spectral points,
has never been observed in nearby blazars, for which the EBL 
absorption effect is negligible.

In order to estimate the redshift
$z^*$ for which the TeV spectral slope equals
the GeV one, the measured spectral points of each source 
have been corrected for the corresponding 
absorption factor starting from $z=0.01$,
and the resulting spectrum fitted with a power law. 
This procedure, applied in fine steps of 
redshift, is iterated until the slope of the 
deabsorbed spectrum equals the one 
measured by LAT. The corresponding redshift, $z^*$, 
is the limit value on the source distance.

\section{Results}
Of the sixteen sources considered in this study,
14 blazars have well known redshift and are used to test
the method, while the remaining  
two blazars (3C~66A and S5~0716+714) have uncertain redshift, 
and are considered separately.
The central columns of Table~\ref{table_values} reports the 
$z^*$ calculated following the method 
described in the previous section, using three different EBL models:
a low limit model~(Kneiske~$\&$~Dole~2010), a mean~(Franceschini~et~al.~2008) 
and a high level one~(Stecker~et~al.~2006, baseline model). 
The absorption coefficients of the last model 
were obtained from a simple extrapolation of the values given for fixed 
redshifts in Stecker~et~al.~(2006) (F. Stecker, private communication).
The errors on $z^*$ are estimated taking into account both 
 errors on the TeV and LAT slopes.
 
\begin{center}
\begin{figure*}
{\small
 \centering
  \includegraphics[width=7.2in]{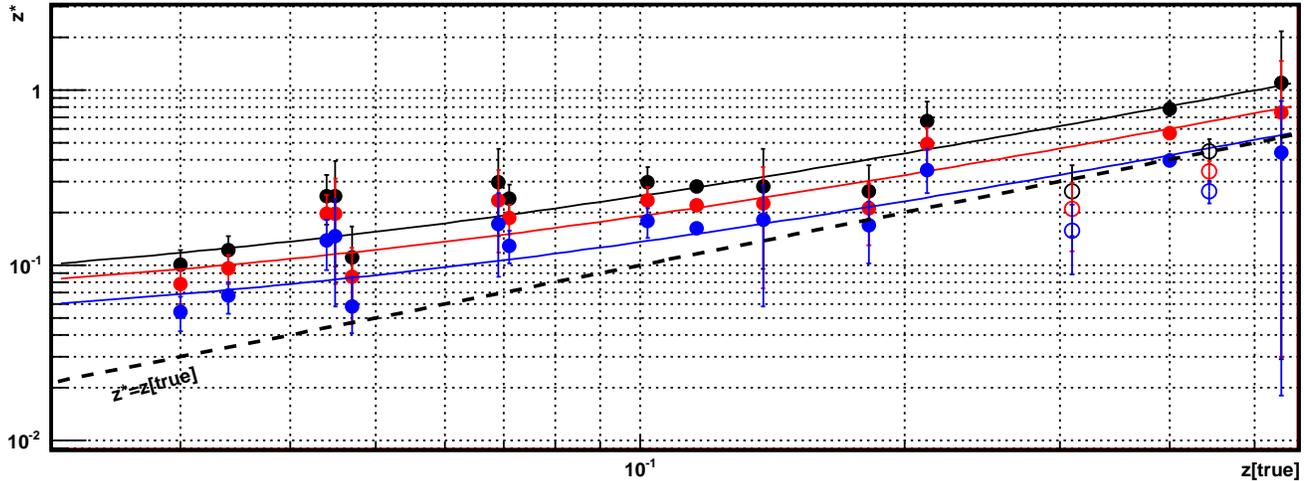}
  \caption{Comparison of the true redshift vs $z^*$ in log scale using three
    different EBL models: black points Kneiske~et~al.~(2010) model (low level), 
    red line Franceschini~et~al.~(2008) model (mean level), blue line 
    Stecker~et~al.~(2006) model (high level). The open points were not used in the fit 
    calculation since their redshift 
    is uncertain (sources 3C~66A and S5~0716+714).}
  \label{correlationplot_comparison}}
\end{figure*}
\end{center}

Fig.~\ref{correlationplot} shows the comparison between 
the real redshift, x-axis,
and the estimated one, y-axis, obtained with the mean 
EBL density model.
All the $z^*$ lie above the bisector (dashed line) 
meaning that their values are larger than the real redshift ones. This 
is expected since we are not considering the presence of the intrinsic break in
the blazar spectra. This result confirms that the method can be used to
 set safe upper
limits on blazars distance. The only exceptions are the two sources 
with uncertain distance, S~0716+714 and 3C~66A (open circles).

Stecker and Scully (2010) derived a linear expression 
for the steepening of the observed TeV slope due to EBL absorption.
Since in our procedure 
$z^*$ is related to this steepening, it is natural to assume that 
also $z^*$ and $z[true]$ are related by a linear function, $z^*=A+Bz[true]$.
The meaning of the coefficients is rather transparent: 
basically $A$ is a measure of the intrinsic 
spectral break of the sources, while, following Stecker $\&$ 
Scully~(2010), $B$ is a measure (increasing values
for decreasing EBL level) of the optical depth of the EBL model used. 

We interpolate with this linear function 
the data with well known distance of Fig.~\ref{correlationplot}.
The linear fit (continuous line) describes very well the data,  
as confirmed by the reduced chi-squared value of the fit,
 $\chi^2/d.o.f.=9.9/12$, and  
corresponding probability of $62\%$.
The results obtained with the other two EBL models are
drawn in Fig.~\ref{correlationplot_comparison}.
It is evident that with a low photon density 
EBL model (black circles),
the estimated redshifts are all shifted at higher values, while with a 
high photon density model (blue circles) the shift
is downwards. Even if the optical depth evolution
is different between the EBL models used here,
the linear behaviour is evident also in the two extreme cases.
The parameters are listed in Table~\ref{params}.
\begin{table}
  \centering
  \begin{tabular}{ l c c}
  \hline
 EBL Model    & $A$ & $B$    \\
 \hline
  \hline
 Low level   & 0.062$\pm$0.017  & 1.86$\pm$0.17  \\
 \hline
 Mean level  & 0.054$\pm$0.012 & 1.36$\pm$0.14 \\ 
 \hline
 High level  & 0.040$\pm$0.009 & 0.96$\pm$0.08  \\
 \hline
\end{tabular}
\caption{Parameters of the linear fitting curves ($z^*$=$A+Bz[true]$) 
plotted in Fig.~\ref{correlationplot_comparison}}\label{params}
\end{table}

Having derived this empirical relation we can try to use it to
{\it derive the redshift} of sources with uncertain distance. This can be done
under the assumption that the source of interest shares similar 
spectral properties with the sources used to derive the fit 
(basically they have similar
values of the spectral break measured by $A$).

\begin{figure}
  \centering
  \includegraphics[width=3.4in]{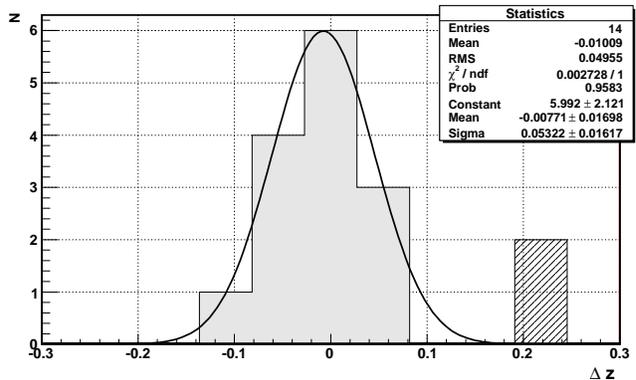}
 \caption{$\Delta$z plot (z[true]-z[rec])
    (filled histogram) and superimposed the two sources with 
    uncertain redshift (S5~0716+714 and 3C~66A), 
    not used for the Gaussian fit.}
  \label{dispersion_plot}
\end{figure}

To demonstrate the feasibility of such a method, in the
last column of Table~\ref{table_values} we report the values of the
 {\it reconstructed} 
redshift, $z[rec]$, obtained by applying the inverse formula 
($z[rec]=(z^*-A)/B$) 
to the $z^*$ estimated with the Franceschini et al. EBL model.
In order to avoid a bias, the parameters $A$ and $B$ are each time calculated
excluding from the fit the source for which we estimate the redshift $z[rec]$.
The differences between the real and the reconstructed redshifts,
$\Delta z$ is drawn in Figure~\ref{dispersion_plot}, filled area. 
Despite the low statistic, the distribution is quite
well described by a Gaussian centered in zero
with a $\sigma$ of $0.05$. The separated shaded
histogram represents the $\Delta z$ of the uncertain redshift sources.

\section{Discussion and Conclusions}

We presented a method that allows the estimation of the quantity $z^*$,
 upper limit on the redshift
of a TeV emitting blazar with a GeV counterpart observed by {\it Fermi}/LAT, 
obtained by deabsorbing the observed TeV spectrum. 

In order to use the largest sample of spectra for this study, 
we made several assumptions: first of all we combined
GeV and TeV data even if the observations in
the different energy bands
were not simultaneous. The impact of this choice,
however, is probably moderated by fact that  
we do not use the flux but only the values of the slopes, 
less variable than the flux (unless in extreme states).
For example the spectral slope of the HBL 1ES~1218+304 
recently measured by the Veritas Collaboration (Acciari~et~al.~2010b)
during a high flux level,
matches within the errors the slope determined during its 
quiescient state. 

Secondly, in this work we use TeV spectra observed with 
various Cherenkov experiments,
characterized by different sensitivities. This difference, 
especially at high energies, 
could affect the result, leading to systematic effects in the distance
limit determination.
Another possible cause of systematics could be the use of all the
blazar sample, independently from the nature of the source: we didn't
apply any distinction between HBL, LBL and FSRQ, characterized 
by a different position of the IC peak.

Despite all these approximations, 
the method presented in this paper applied to a 
sample of test sources gave satisfactory results.
The $z^*$ values obtained by correcting the spectra
from the EBL absorption, are, in fact, 
all above the real redshift values if we use a mean
background photon level. This suggest the use
of this method for constraining the distance of 
unknown redshift sources.

We applied the $z^*$ estimate also to two sources with
uncertain distance: in both cases the 
limit lies below the quoted values. This result could be
due to some intrinsic properties of the sources (specifically,
a more moderate intrinsic spectral break between the GeV and TeV bands than 
that of the other sources), or 
to a wrong estimate of their distances. 
In the latter case, our method would constrain the redshift of
S5~0716+714 below $0.21\pm0.09$ and that of 3C~66A 
below $0.34\pm0.05$. 
It can be pointed out that in the case of S5~0716+714,
the redshift of 0.31 used in this work, recently reported
by Nilsson et al. (2008), is estimated by assuming the luminosity
of its host galaxy. 
Another estimate on the blazar distance, based on the 
spectrography of the three galaxies close to this source, 
gives the value of $\sim 0.26$ for its redshift, more in
agreement with our derived limit.

The same procedure was applied to our sample using two extreme EBL
models. The low density one gives even safer 
upper limits on the sources distances,
while the $z^*$ obtained with the high density model
are closer to the real redshift values.
Even with this model, all the estimates are above or on the bisector,
confirming our assumption that the deabsorbed TeV slope cannot be
harder than the GeV slope reported by {\it Fermi}/LAT.

Following previous works, we tested the
possibility of a linear relation between our 
$z^*$ estimates and the real distances of the sources.
We found that the linear fit describes quite well our results, independently 
on the EBL model considered, although the slope and
intercept of the fits are different in the three cases
(Table~\ref{params}). 

The relation found suggests to use the $z^*$ estimate
not only to set an upper limit on unknown distances
of blazars, but also, via the inverse-formula, to try an
evaluation of this distance. In order to
investigate this opportunity, we tested it on our sample
of sources using the mean EBL model, paying a 
special attention to avoid biases in the calculation.
The distribution of the difference $\Delta z$  
between the reconstructed and the real
redshift is well described by a Gaussian peaked in zero
with a $\sigma$ of $0.05$. Once again, the uncertain 
redshift sources are outside the
expected interval.
The value of the redshift of S5~0716+714 obtained with this
method is $0.11\pm0.05$, where the error quoted is
the $\sigma$ of the $\Delta z$ distribution. For
3C~66A, the same procedure leads to a redshift
estimate of $0.213\pm0.05$. 

As a final example of application, we use our procedure 
to PKS~1424+240, a blazar of unknown redshift
recently observed in the VHE regime by Veritas~(Acciari~et~al.~2010a). 
The {\it Fermi}/LAT spectrum slope measured between $0.2$ to $300$ GeV is 
$1.85\pm0.05$ and the corresponding $z^*$
redshift at which the deabsorbed TeV spectrum slope
equals it is $0.382\pm0.105$, using Franceschini et al. EBL model.
This result is in agreement with the value of $0.5\pm0.1$, 
reported by Acciari~et~al.~(2010a),
calculated applying the same procedure 
but only simultaneous {\it Fermi} data.
Our estimate on the most probable distance for PKS~1424+240,
obtained by inverting the $z^*$ formula,
is $z[rec]=0.24\pm0.05$, where the
error, as before, is assumed as the $\sigma$ of the Gaussian 
fitting the $\Delta z$ of Fig.~\ref{dispersion_plot}. 

\section*{ACKNOWLEDGMENTS}
We are grateful to A.~Franceschini, T.~M.~Kneiske and F.~W.~Stecker 
for providing the optical depth data and C.~W.~Danforth for
the communication of the new determination of PG~1553+113 redshift.
GB, LM and FT acknowledge a 2007 Prin-MIUR grant for financial support.

\label{lastpage}

\end{document}